\RequirePackage[2020-02-02]{latexrelease}
\documentclass[twocolumn,amsmath,amssymb]{revtex4}

\usepackage{graphicx}
\usepackage{dcolumn}
\usepackage{bm}
\usepackage{afterpage}
\usepackage{placeins}
\usepackage{ragged2e}
\usepackage{amsmath}
\usepackage{lipsum}

\usepackage{nccmath}
\usepackage{amsmath}
\begin{document}


\title{ 
	Static negative susceptibility in ferromagnetic material induced by domain wall motion: an aspect of superconductor state}

\author{Nilesh Pandey}
  \email{pandeyn@iitk.ac.in;chauhan@iitk.ac.in}
 \author{Yogesh Singh Chauhan}
\affiliation{  
Department of Electrical Engineering, Indian Institute of Technology - Kanpur,  Kanpur, 208016, India
}
	
\begin{abstract}
Domain wall motion in magnetic materiel induces the negative susceptibility leading to a perfect diamagnetism state. The local susceptibility is calculated by the derivative of magnetization ($\vec{M}$) vector w.r.t. magnetic field strength ($\vec{H}$) vector. In the transient region from the upward domain to the downward domain (domain wall width), local $\vec{M}$ and $\vec{H}$ vectors exhibit opposite slopes, which leads to a negative susceptibility value. 
A negative susceptibility value induces the diamagnetism effect leading to a relative permeability value $<$ 1 $\left(\mu _r < 1\right)$.  This diamagnetism sate originates due to the domain wall motion, which is an entirely different mechanism from the electron motion's induced diamagnetism. Furthermore, the strength of the diamagnetism state can be enhanced by tuning the gradient energy of a domain that may correspond to a perfect diamagnetism state $\left(\chi _v \approx -1 \Rightarrow \mu _r \rightarrow 0\right)$. Besides, we believe that there may be a possibility of sustaining such a diamagnetic state (domain wall induced) in a ferromagnetic material that is utterly contradictory to the conventional theory.

\end{abstract}

\maketitle
\section {Introduction}
Diamagnetism is a well-known phenomenon that has been studied for many decades \cite{Langevin}-\cite{Cullity}. It is demonstrated that diamagnetism is a material property caused by the electron's motion and spins\cite{Langevin}-\cite{Cullity}. In general, diamagnetism is a weak force that leads to a small negative susceptibility value. Hence, the relative permeability of diamagnetic materials is close to unity \cite{Jiles}-\cite{Cullity}.

In the literature, both classical and quantum theory of diamagnetism are focused on the electron's orbital motions and magnetic moment \cite{Langevin}-\cite{Cullity}. However, the domain theory of diamagnetism is still not reported! 
This article analyzes the local spatial distribution of  magnetization ($\vec{M}$) and magnetic field ($\vec{H}$) vectors.
Subsequently, local susceptibility is calculated by the derivative of $\vec{M}$ w.r.t. $\vec{H}$. We observed that local susceptibility could attain a negative value in the domain wall transition region, which induces the diamagnetism effect. Furthermore, local diamagnetism varies with the domain's gradient energy that can be tuned to achieve a superconducting state $\left(\mu_r \sim 0\right)$. \vspace{-5mm}
\section{Problem statement and methodology}
The local relative permeability is defined as.
\begin{ceqn}
\begin{align}
	\mu_r=1+\frac{\partial M }{\partial H}
\end{align} 
Net magnetic field is calculated by the vector sum of an external field $\left(H_{ext}\right)$ and a demagnetized field $\left(H_{int}\right)$ \cite{Jiles}.
\begin{align}
	\vec{H}=\vec{H_{ext}}+\vec{H_{int}} \label{H_field}
\end{align}  
The local volume susceptibility is defined as.
\begin{align}
\chi_v=\frac{\chi_e\chi_i}{\chi_e+\chi_i} \label{kappa}
\end{align}
where, $\chi_e$ is due to the external magnetic field, and  $\chi_i$ is due to the internal demagnetization field.
Note that $\chi_e$ is always positive, and $\chi_i$ is a negative quantity. Thus, the condition for $\chi_v < 0$ is derived as.
\begin{align}\nonumber
\chi_i+\chi_e > 0 \\ \nonumber
\Rightarrow  \chi_i > - \chi_e \\ 
\Rightarrow \left | \chi_i \right | < \left | \chi_e \right | \label{condn_1}
\end{align}
Net volume susceptibility from $\left(\ref{kappa}\right)$ is a parallel combination of $\chi_e$ and $\chi_i$, and for a ferromagnetic material:
\begin{align}
	\left | \chi_e \right | >> 1
\end{align}
Therefore, as long the condition of  $\left | \chi_i \right |<< \left | \chi_e \right |$ is satisfied, the net volume susceptibility can be expressed as.
\begin{align}
	&\frac{1}{\chi_v}=\frac{1}{\chi_e}+\frac{1}{\chi_i}\approx\frac{1}{\chi_i} \label{condn_2} \\ 
	&\Rightarrow \mu_r \sim 1 + \chi_i \label{permbi}
\end{align}
Hence, $\left(\ref{condn_1}\right)$ and $\left(\ref{condn_2}\right)$ are the conditions for a diamagnetic state in the ferromagnetic material. The next task is to derive an accurate expression of the demagnetization field, which will calculate the local susceptibility.
 \begin{figure*}[!t]
	\centering
	\includegraphics[width=1\textwidth]{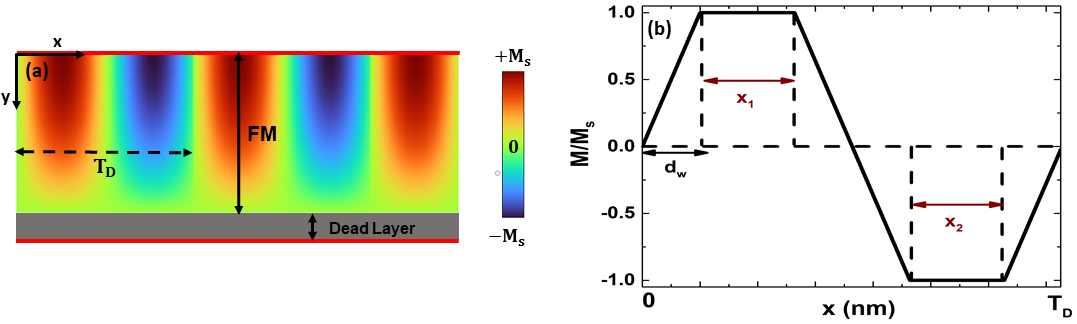}
	\caption{(a) Schematic of a multi-domain ferromagnetic stacked layer with the dead layer. (b) Schematic of magnetization domain wave with domain wall width. Default material parameters are taken from \cite{Jiles}.
}\label{fig:schm}
\end{figure*}

\subsection{Magnetostatics model development}
Fig. \ref{fig:schm}(a) shows the schematic of a ferromagnetic multi-domain layer with an interfacial dead layer. 
The motivation to include the dead layer is two-fold: studying the impact of the domain's gradient energy on diamagnetism and applying the smooth Dirichlet and Neumann boundary conditions at the interfaces.\cite{Jackson}.
Additionally, numerous experimental demonstrations of the ferromagnetic layer with a dead layer were also reported in the literature \cite{Huijben}-\cite{Yasuda}. Note that the developed physics and model in this work is also valid for a device without a dead layer. 
\\

The magnetostatics 2-D Poisson's equation with effective magnetic charge density is expressed as \cite{Jackson}.
\begin{align}
&	\frac{\partial ^2 \phi_{fm}}{\partial x^2}+\frac{\partial ^2 \phi_{fm}}{\partial y^2}=\frac{\partial M}{\partial x}+\frac{\partial M}{\partial y}  \left(\text{Ferromagnetic layer}\right) \label{Poisson} \\
	&	\frac{\partial ^2 \phi_{ox}}{\partial x^2}+\frac{\partial ^2 \phi_{ox}}{\partial y^2}=0 \left(\text{Dead layer}\right)\label{DE}
\end{align}

The magnetization vector is assumed to follow a periodic nature, which is a valid approximation for a strip domain structure \cite{Jiles}. Therefore, the mathematical formulation of the magnetization vector can be obtained in the Fourier series.
\begin{align}\nonumber
	&M(x,y)=\\ 
	&M_sf(y)\left(\frac{a_0}{2}+\frac{2}{d_wT_D} \sum_{j}\left \{\frac{ A_jcos\left ( k_jx \right )+B_jsin\left ( k_jx \right )}{k_j^2} \right \}\right) \label{pol_wave}
\end{align}
where, $f(y)$ is a unit-less function which is used to incorporate the gradient in a domain magnetization along the vertical direction, $M_s$ is the saturation magnetization, $d_w$ is the domain wall width, $T_D$ is the net domain period, and $a_0$ is the Fourier series coefficient corresponding to $j=0$. We have considered a finite domain wall width $d_w$, which is a significant aspect of capturing the local negative susceptibility. The component $\nabla .M$ signifies the effective magnetic charge density. Both lateral $\left(\partial M/ \partial x\right)$ and vertical $\left(\partial M/ \partial y\right)$ directional gradients are considered in the model. 

Eq. $\left(\ref{Poisson}\right)$ and $\left(\ref{DE}\right)$ possess Dirichlet and Neumann boundary conditions at the boundary interfaces. Solutions of such 2-D Poisson's equation are obtained in the Fourier series form using the Green's function approach \cite{Jackson}, \cite{Lin}-\cite{nilesh}.  The boundary conditions and Fourier coefficients of $\left(\ref{Poisson}\right)$ and $\left(\ref{DE}\right)$ are given in the Appendix section. The derived 2-D distribution of magnetic scalar potential is provided by $\left(\ref{MD_pot}\right)$ and $\left(\ref{ox_plot}\right)$.  


	\noindent {Ferromagnetic layer:}  
	\begin{align}\nonumber
		&\phi_{fm}(x,y)= \phi_{MD}-\frac{2}{L}\sum_{m}\frac{sin\left ( k_{m}x \right )B_{s}^msinh( k_{m}y)}{\mu _{fe} k_{m}cos( k_{m}t_{fm})}\\ \nonumber
		&+\sum_{m}\frac{2sin\left ( k_{m}x \right )cosh( k_{m}\left (t_{fm}-y  \right ))H_0\left ( 1+(-1)^{m+1} \right )}{ k_{m}cos( k_{m}t_{fm})}\\ 
		&+\frac{2}{t_{fm}} \sum_{n}\frac{sin\left (k_{n}^{I}y  \right )\left ( A_{1}^{n}sinh\left ( k_{n}^{I}\left ( L-x \right ) \right )+A_{2}^{n}sinh\left ( k_{n}^{I}x \right ) \right ) }{ sinh( k_{n}^{I}L)} \label{MD_pot}
	\end{align} 
		\noindent{Dead layer:}  
	\begin{align}\nonumber
		&\phi_{ox}(x,y)=\frac{2}{L}\sum_{m}\frac{sin\left ( k_{m}x \right )B_{s}^msinh( k_{m}\left(t_{fm}+t_{ox}-y\right))}{\mu _{ox} k_{m}cos( k_{m}t_{ox})}\\ \nonumber
		&+\sum_{m}\frac{2sin\left ( k_{m}x \right )cosh( k_{m}\left (t_{fm}-y  \right ))H_0\left ( 1+(-1)^{m+1} \right )}{ k_{m}cos( k_{m}t_{ox})}\\ \nonumber
		&+\frac{2}{t_{ox}} \sum_{n}\frac{sin\left (k_{n}^{II}\left(t_{fm}+t_{ox}-y\right)  \right )} { sinh( k_{n}^{II}L)} \\ &\times \left ( B_{1}^{n}sinh\left ( k_{n}^{II}\left ( L-x \right ) \right )+B_{2}^{n}sinh\left (k_{n}^{II}x \right ) \right ) \label{ox_plot}
	\end{align}
\subsection{Domain period calculation}
The formation of domains in a ferromagnetic layer is well studied, and it is demonstrated that the principal cause of domain formation is to minimize net system energy \cite{Jiles}, \cite{Cullity}, \cite{Lemesh}. Therefore, the state and period of the domain in a ferromagnetic layer are evaluated by minimizing the net system energy, leading to a thermodynamically stable state.
The net energy density of the ferromagnetic layer is defined as \cite{Jiles}, \cite{Lemesh}.

\begin{align}
	f_{net}=f_{exch}+f_{anis}+f_{dem}+f_{DMI} \label{f_net}
\end{align}
Interactions between magnetic moments lead to exchange energy density $\left(f_{exch}\right)$ \cite{Jiles},\cite{Lemesh}.
\begin{align}
f_{exch}=A\left\{\left(\frac{\partial m_x}{dx}\right)^2+\left(\frac{\partial m_y}{dx}\right)^2\right\}
\end{align}
Crystal anisotropy has a negligible impact on domain texture \cite{Jiles}. However, we consider the anisotropy effect in the domain formation.
\begin{align}
f_{anis}=k_u\left(	m_x^2+m_y^2\right)
\end{align} 

The demagnetizing fields primarily determine domain growth \cite{Jiles},\cite{Cullity},\cite{Lemesh}.Therefore, demagnetizing energy density is the most significant factor that should be considered to evaluate the domain state.

\begin{align}
	f_{dem}=\frac{\mu _0}{2}M.H=\frac{\mu _0}{2}M.\left(-\nabla  \phi_{fm}\right)
\end{align}

 $k_u$ is the magnetocrystalline anisotropy, $A$ is the exchange stiffness, and $m_x$, $m_y$ are the normalized magnetization $\left(M/M_s\right)$. Angel dependency in the domain wall can be included as.
\begin{align}
	\frac{m_x}{m_y}=tan\left(\theta\right)
\end{align}  

here, $\theta$ is the wall angle: $\theta=0\rightarrow$ Bloch type walls and  $\theta=\pi /2\rightarrow$ Néel type walls.
Dzyaloshinskii-Moriya interaction (DMI) energy density is defined as \cite{Lemesh}.
\begin{align}
f_{DMI}=\pi D sin\left(\theta\right)
\end{align}

The state of a domain can be completely determined by the net energy density given in $\left(\ref{f_net}\right)$ \cite{Jiles},\cite{Cullity}.
The default values of ferromagnetic material parameters are taken from \cite{Jiles}.  

Net energy is calculated by the volume integral of energy density.
\begin{align}
	F_{net}=W\int_{0}^{L}\int_{0}^{t_{fm}}f_{net}dydx \label{F_Net}
\end{align}
where, $W$ is the width of device = 1 $\mu$m. Eq. $\left(\ref{F_Net}\right)$ is differentiated w.r.t domain period ($T_D$), domain wall width ($d_w$), and domain wall angle ($\theta$), forming a system of equations. This system of equations is equated to 0 to obtain the state parameters of the domain ($T_D$, $d_w$, and $\theta$ ). 
\begin{figure}[!t]
	\centering
	\includegraphics[width=0.5\textwidth]{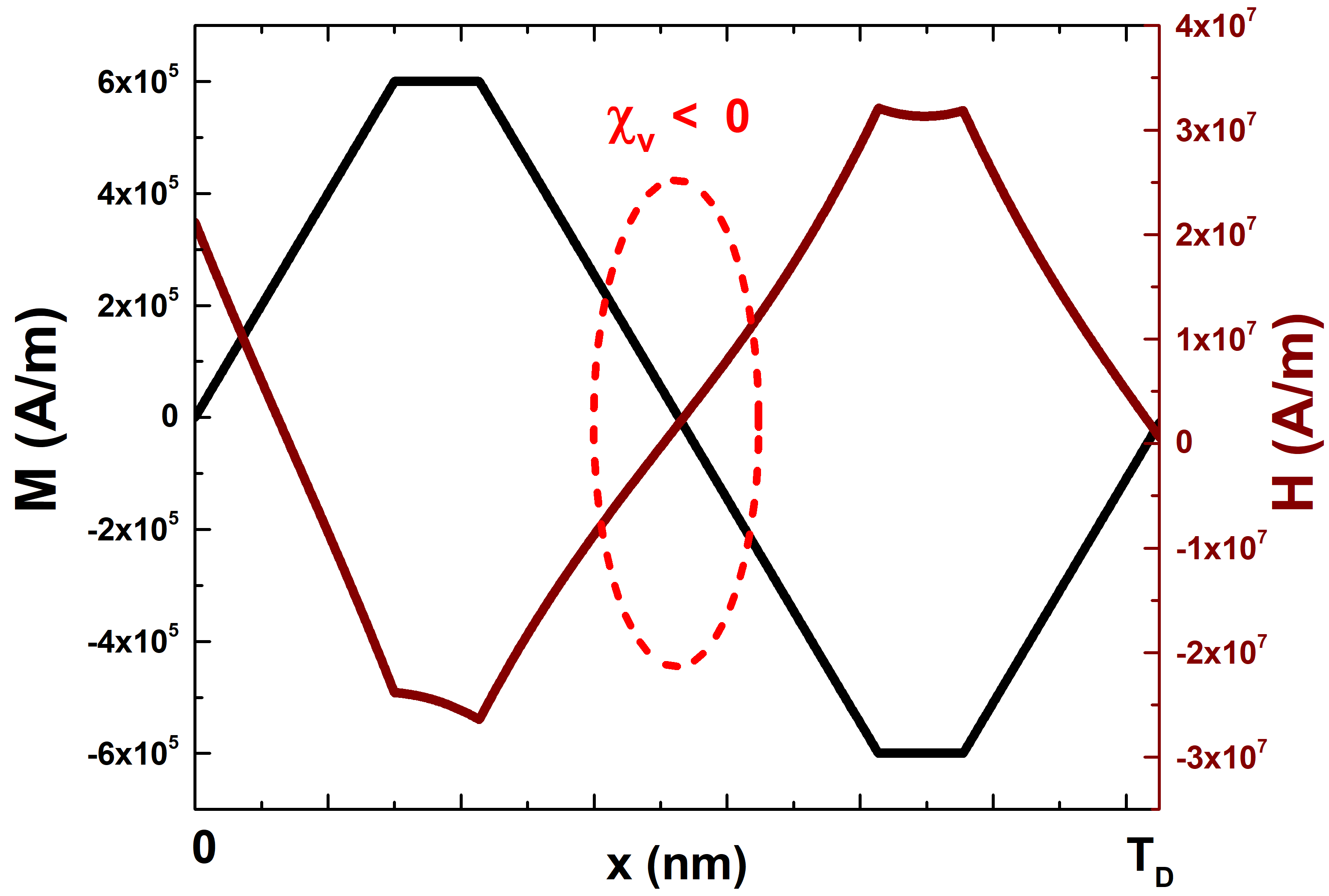}
	\caption{Signature of diamagnetism via domain wall motion. 
	}\label{fig:Xv_1}
\end{figure}
\begin{figure}[!b]
	\centering
	\includegraphics[width=0.5\textwidth]{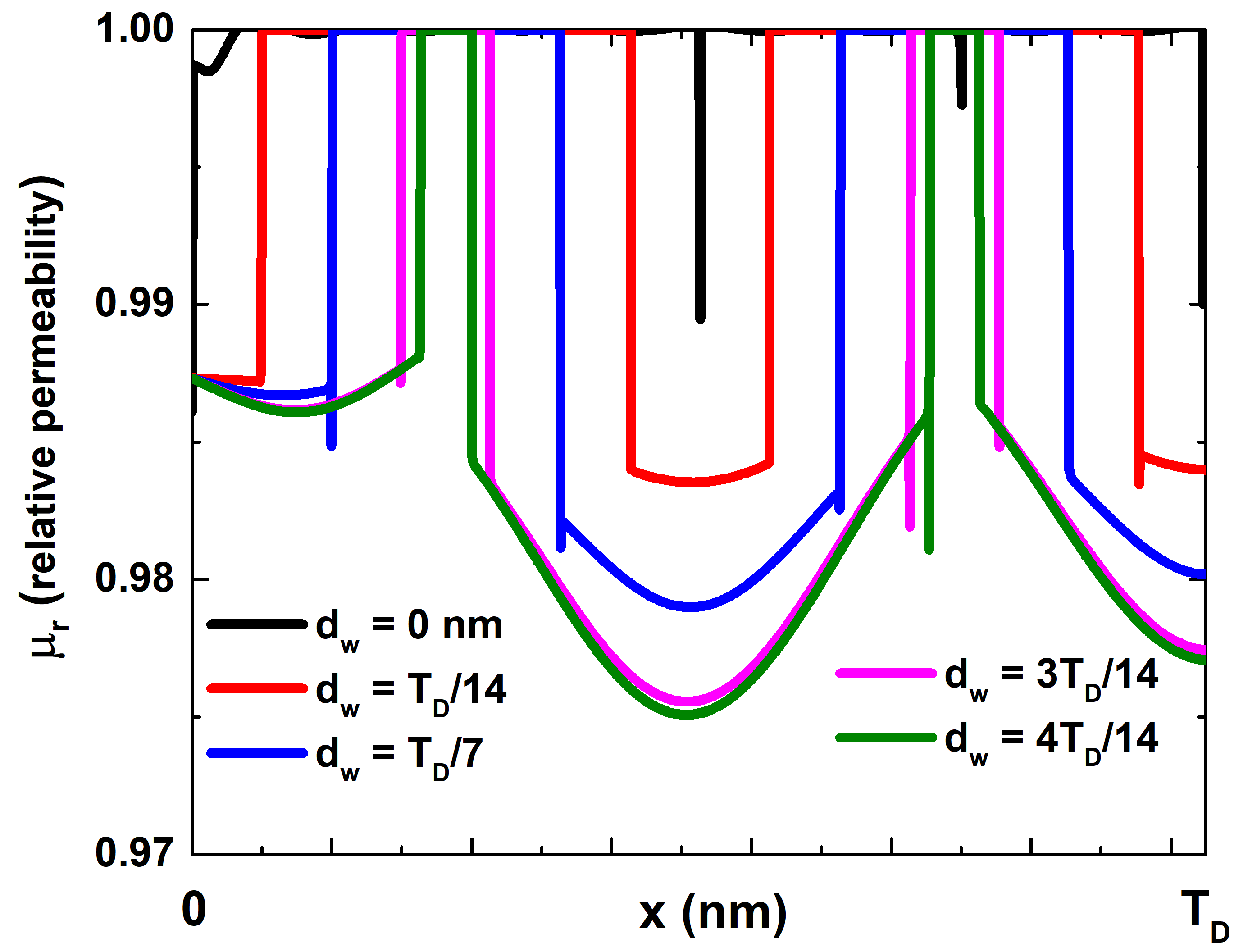}
	\caption{Local relative permeability for the various domain wall width, at $y=t_{fm}/2$ (without vertical directional gradient in domain). 
	}\label{fig:ur_w_o_grad}
\end{figure}
\section{Results and Discussion}
\subsection{Domain wall induced diamagnetism}
Fig. \ref{fig:Xv_1} shows the local distribution of magnetization $\vec{M}$ and $\vec{H}$. It is remarkable to observe that $\vec{M}$ and $\vec{H}$ exhibit an opposite slope in the domain wall region, which induces the diamagnetism effect. Local susceptibility of the material is defined as.

\begin{align}
	\chi_i (x)=\frac{\partial M}{\partial H }=\frac{\partial M}{\partial x }\left(\frac{\partial H}{\partial x }\right)^{-1}
	\label{local_subs}
\end{align}
It is clear from Fig. \ref{fig:Xv_1} that $\partial M/\partial x < 0$ and $\partial H/\partial x > 0$. Therefore, in the domain wall region, local susceptibility (\ref{local_subs}) will be a negative quantity which is a necessary condition for diamagnetism \cite{Jiles},\cite{Cullity}. 
\begin{figure}[!t]
	\centering
	\includegraphics[width=0.5\textwidth]{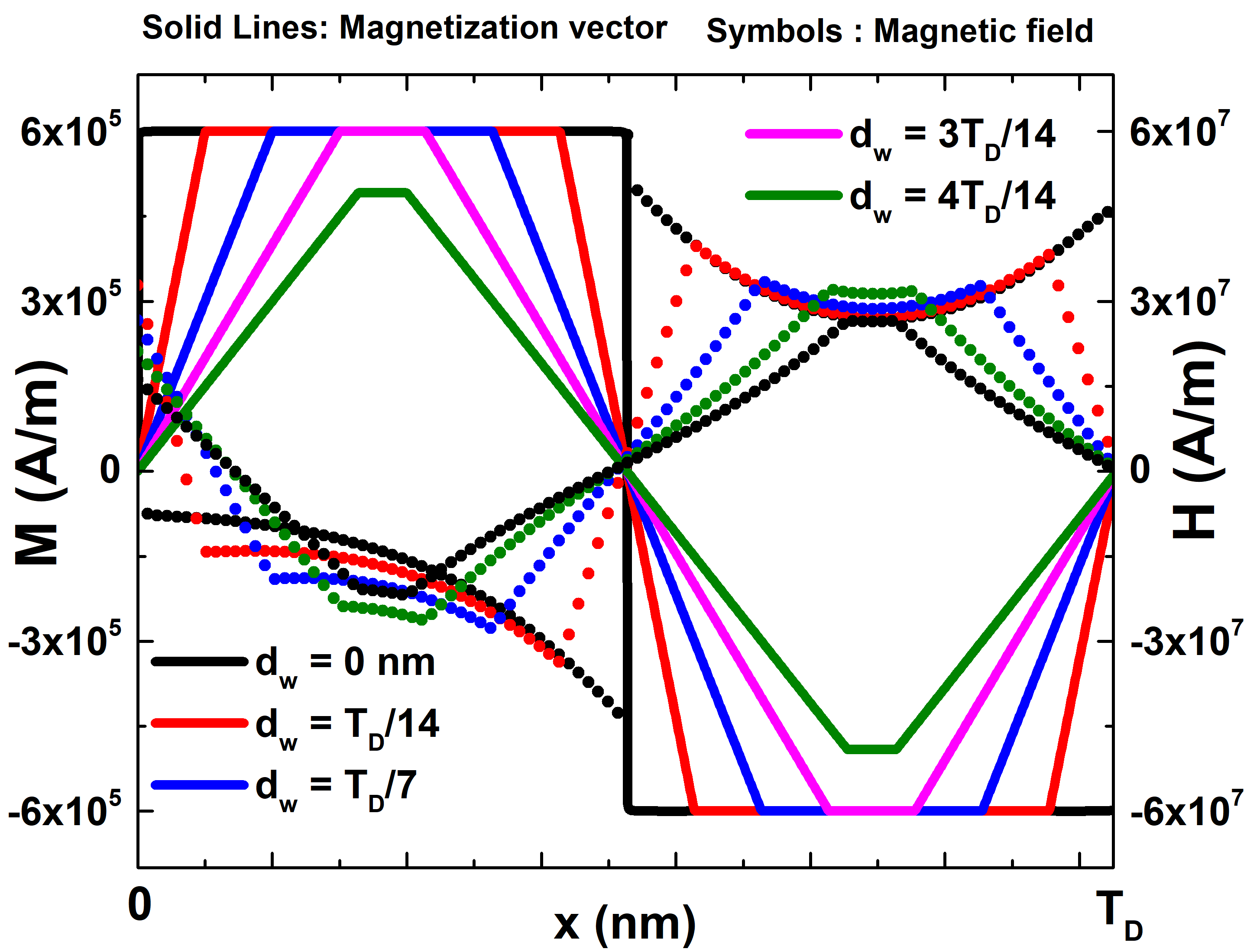}
	\caption{Local distribution of $\vec{M}$ and $\vec{H}$ for the various domain wall width at $y=t_{fm}/2$ (without vertical directional gradient in domain). 
	}\label{fig:M_H_dw}
\end{figure}
\begin{figure}[!b] \hspace{-7mm}
	\centering 
	\includegraphics[width=0.5\textwidth]{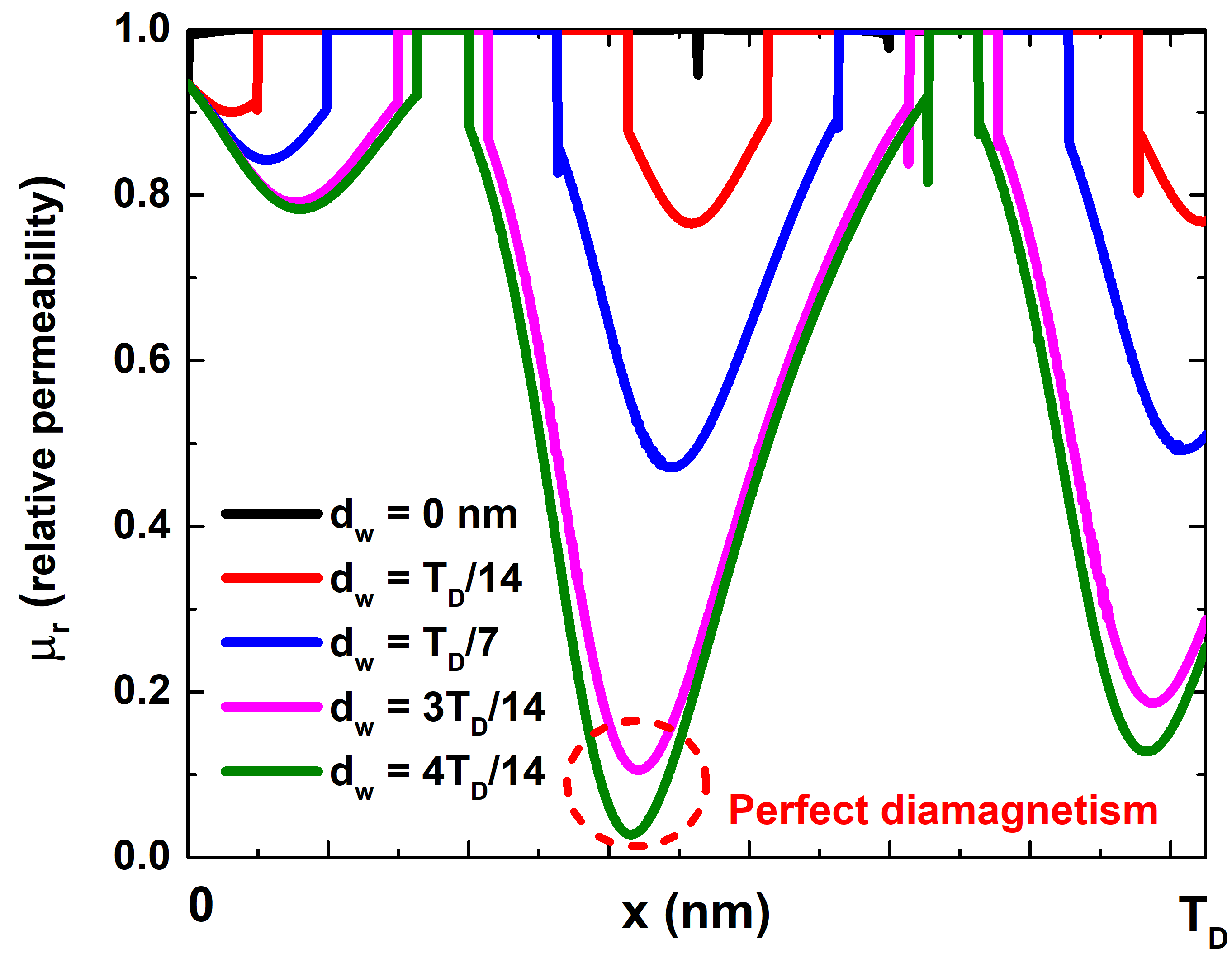}
	\caption{Local distribution of relative permeability plotted at y $\sim 0$, with the gradient in domain along vertical direction. Possibility of perfect diamagnetism! an aspect of superconductor!  
	}\label{fig:ur_grad_tfe_50}
\end{figure}
Fig. \ref{fig:ur_w_o_grad} shows the local relative permeability $\left(1+\chi_i (x)\right)$ plot for the various domain wall widths. Due to a negative value of susceptibility, the value of $\mu _r$ is always less than unity, thus proving the presence of diamagnetism effect in domain wall width. The strength of diamagnetism enhances with the increasing value of the domain wall. To analyze the dependency of diamagnetism on domain wall width, Fig. \ref{fig:M_H_dw} is shown. An increase in the $d_w$ raises the local spatial slope (w.r.t $x$) of $\vec{M}$ and $\vec{H}$. However, the magnitude of relative slopes (\ref{local_subs}) increases, which enhances the diamagnetism effect. Hence, $\mu_r$ decreases with the domain period. The minimum value of $\chi_i$ is $\sim$ -2.49$\times$10$^{-2}$, which shows a significantly higher diamagnetism than the conventional diamagnetic materials that have $\chi$ in the range of 10$^{-5}-$10$^{-4}$ \cite{Jiles}.

Therefore, domain wall-induced diamagnetism is a strong force that can exist in the ferromagnetic substance. However, the minimum value of $\mu_r$ is $\approx$ 0.975, which is still far more significant than the perfect diamagnetism state. $\left(\mu_r \sim 0\right)$. 

\subsection{Possibility for perfect diamagnetism}
The strength of domain wall-induced diamagnetism can be further enhanced by introducing a gradient in the magnetic domain along the vertical direction. Fig. \ref{fig:schm}(a) shows the ferromagnetic-dead layer structure. Since the dead layer does not have any ferromagnetic domain, the magnetization vector must be zero at the ferromagnetic/dead layer interface (assuming zero surface current density at the interface). The boundary conditions at the interface are given in $\left(\ref{BCs}\right)$. The function $f(y)$ is used to incorporate the vertical directional gradient is defined by $\left(\ref{f_y}\right)$. From $\left(\ref{f_y}\right)$, it can be observed that the magnetization vector is maximum at $y=0$ and minimum at the ferromagnetic/dead layer interface. Similar kind of observations were also reported for the ferroelectric-dead layer interface \cite{Park}, \cite{Saha_3}, \cite{nilesh_1}. 
\begin{align} \nonumber
	&	\phi_{fm}(x,t_{fm})=\phi_{ox}(x,t_{fm}) \\ \nonumber
	&\Rightarrow \frac{\partial \phi_{fm}(x,t_{fm})}{dx}=\frac{\partial \phi_{ox}(x,t_{fm})}{dx}  \\ \nonumber
	&\text{Since, }\; M =4 \pi \nabla\phi 
	\\ \nonumber
	&	\Rightarrow M_{fm}(x,t_{fm})=M_{ox}(x,t_{fm})\; \\ \nonumber
	&\Rightarrow H_{1,t}=H_{2,t}\;\left(\text{at zero surface current.}\right) \\ \nonumber
	& M_{ox}(x,t_{fm}) = 0\;\;\;\; \left(\text{no domains in the dead layer}\right)
	\\ 
	& \Rightarrow M_{fm}(x,t_{fm})=0 \label{BCs}
	\\
	& f(y)=\left(\frac{-1}{t_{fm}}\right)^2y+1 \label{f_y}
\end{align}

Fig. \ref{fig:ur_grad_tfe_50} shows the local permeability profile plotted along with the $y=0$ interface. 
A gradient in the magnetization vector shows a remarkable result. Now, the local value of $\chi_i$ can approach -1, leading to a perfect diamagnetism state $\left(\mu_r \sim 0\right)$. This can be explained by observing the local spatial slope of $\vec{M}$ and $\vec{H}$. At $y=0$, spatial $x$ directional variations in $\vec{M}$ are huge which is evident in Fig. \ref{fig:schm}(a). The $\vec{M}$ oscillates between $+M_s$ to $-M_s$ $\left(\Delta M_s = 2\left | M_s\right |\right)$ in a very narrow range of domain wall width ($\sim$ few nm), leading to an enormously larger slope of $\vec{M}$ $\left(\partial M/\partial x\; \text{is very large}\right)$. On the other hand, the spatial variation $\vec{H}$ w.r.t $x$ is also present. But the fractional change in $\vec{M}$ dominates over fractional change in $\vec{H}$ $\left(\partial M/\partial x>> \partial H/\partial x\right)$. Therefore, at some specific points, local permeability approaches zero! At these specific points, $\left | \chi_i \right |$ is the maximum these $x$ values can be obtained as.
 
\begin{align}
	\frac{\partial \chi_i }{\partial x}=0
\end{align}

Fig. \ref{fig:ur_grad_tfe_2} is plotted at $y=t_{fm}/2$. The diamagnetism strength decreases due to a reduction in the $\partial M/\partial x$ component (see Fig. \ref{fig:schm}(a)). An increase in the value of $y$ decreases the value of $f(y)$, which reduces the gradient along the y-direction. 
Hence, oscillations in the magnetization vector decrease $\left(\Delta M_s < 2\left |  M_s\right |\right)$, reducing the diamagnetism strength. Therefore, the minimum value of $\mu_r \sim $ 0.89.
At the ferromagnetic-dead layer interface (y=$t_{fm}$), $\vec{M}$ spatial $x$ directional variations are negligible compared to spatial variations at y=0 $\left(  \partial M/\partial x |_{y=t_{fm}}<<\partial M/\partial x |_{y=0}\right)$.

\begin{figure}[!t]
	\centering
	\includegraphics[width=0.5\textwidth]{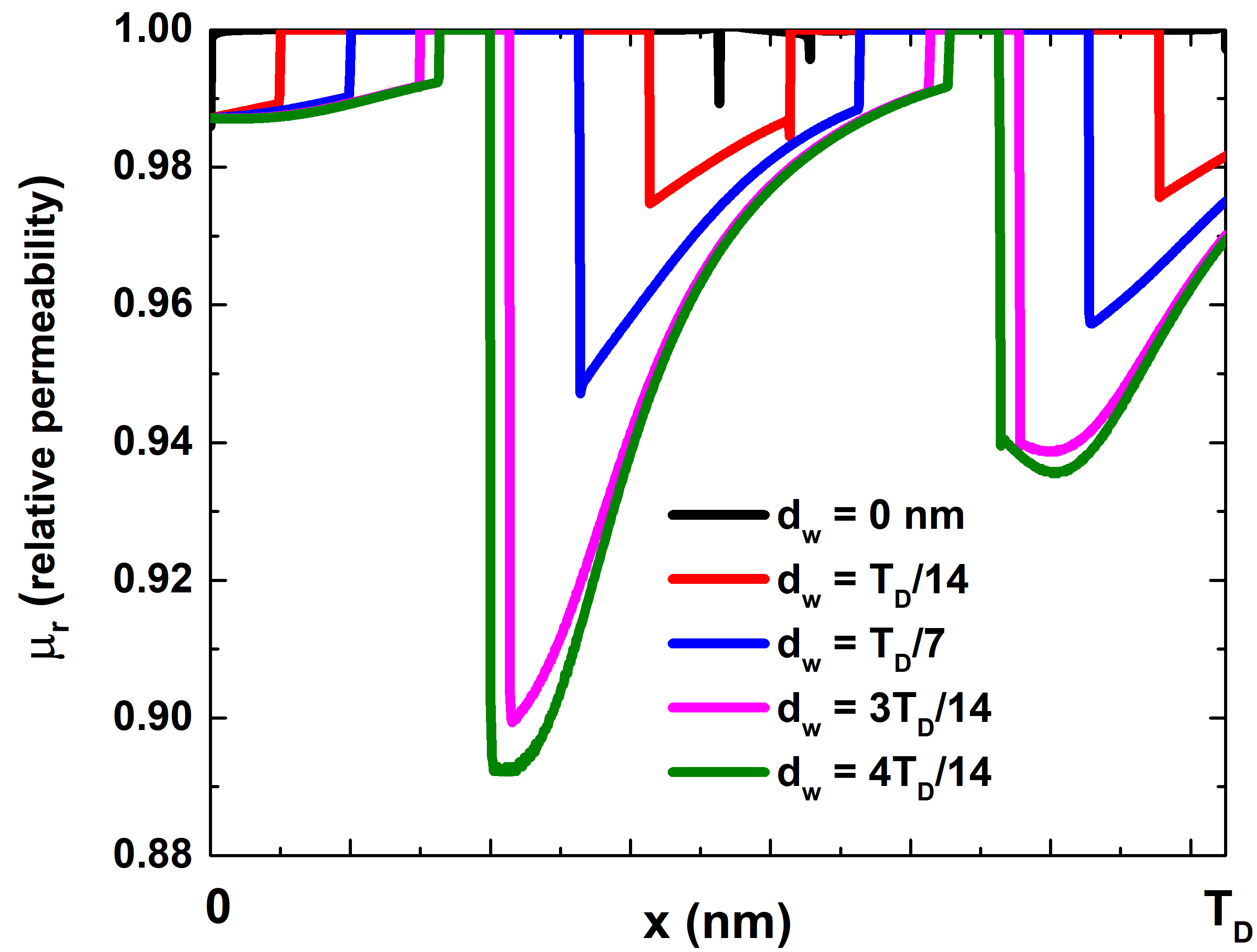}
	\caption{Local distribution of relative permeability plotted at y= $t_{fm}/2$, with the gradient in domain along vertical direction. 
	}\label{fig:ur_grad_tfe_2}
\end{figure} 

\section{conclusion}
We have studied the possibility of attaining a perfect diamagnetism via domain wall motion. In the domain wall width region, 
magnetization and magnetic field vector exhibit opposite slopes, leading to a negative susceptibility value. The origin of negative susceptibility is a macroscopic and static phenomenon entirely different from the conventional diamagnetism caused by the electron's spin and momentum. The strength of domain wall-induced diamagnetism can be increased drastically by stacking a ferromagnetic layer with the dead layer. The gradient in the domain (caused by the dead layer) raises the effectiveness of diamagnetism. Therefore, perfect diamagnetism can be obtained by tuning the gradient energy of a domain. At the ideal diamagnetism state, a material's local relative permeability can attain a zero value corresponding to an aspect of the superconductor.

The experimental observations of such domain wall-induced diamagnetism are tedious and can be explained by analyzing the local spatial distribution of susceptibility. The state of perfect diamagnetism exists only for a small nanoscale range (see Fig. \ref{fig:ur_grad_tfe_50}) that could be challenging to measure such nano-scaled variations in the device. Furthermore, domain wall dynamics vary significantly by an external applied magnetic field, which may destroy the state of perfect diamagnetism. Therefore, it can be stated that this work formulated a theoretical study of domain wall-induced diamagnetism that can exist in the ferromagnetic material! However, experimental demonstration of such domain wall-induced diamagnetism will require a through analysis of the local distribution of $\vec{M}$ and $\vec{H}$.

\section*{APPENDIX}
Green's functions for ferromagnetic and dead layer regions.\\

\noindent I. Ferroelectric region:
\begin{align}\nonumber
&G^{fe}_x(x,y;x',y')= \\ \nonumber
&\frac{2}{L}\sum_{m=1}^{\infty}\frac{sin\left(k_mx\right)sin\left(k_mx'\right)}{k_mcosh\left(k_mt_{fe}\right)} \times\\ \nonumber
&\big[sinh\left(k_my\right)cosh\left(k_m\left(t_{fe}-y'\right)\right)\;\;\;\;\; y<y' \\ \nonumber
&sinh\left(k_my'\right)cosh\left(k_m\left(t_{fe}-y\right)\right)\;\;\;\;\;\; y>y'\big]
\\ \\\nonumber
&G^{fe}_y(x,y;x',y')= \\ \nonumber
&\frac{2}{t_{fe}}\sum_{n=1}^{\infty}\frac{sin\left(k_n^Iy\right)sin\left(k_n^Iy'\right)}{k_n^Isinh\left(k_n^IL\right)}\times\;\;\;\;\;\;\;\;\;\;\;\;\;\;\;\;\;\;\;\;\;\;\;\;\;\;\;\;\;\; \\ \nonumber
&\big[sinh\left(k_n^Ix\right)sinh\left(k_n^I\left(L-x'\right)\right)\;\;\;\;\; x<x' \\ 
&sinh\left(k_n^Ix'\right)sinh\left(k_n^I\left(L-x\right)\right)\;\;\;\;\; x>x\big]
\end{align}
II. Dead layer:
\begin{align}\nonumber
&G^{ox}_x(x,y;x',y')= \\ \nonumber
&\frac{2}{L}\sum_{m=1}^{\infty}\frac{sin\left(k_mx\right)sin\left(k_mx'\right)}{k_mcosh\left(k_mt_{ox}\right)}\times\;\;\;\;\;\;\;\;\;\;\;\;\;\;\;\;\;\;\;\;\;\;\;\;\;\;\;\;\;\;\\ \nonumber
&\big[cosh\left(k_m\left(t_{fe}-y\right)\right)sinh\left(k_m\left(t_{1}-y'\right)\right)\;\;\;\;\; y<y' \\ 
&cosh\left(k_m\left(t_{fe}-y'\right)\right)sinh\left(k_m\left(t_{1}-y\right)\right)\;\;\;\;\;\; y>y'\big]
\end{align}
\begin{align}\nonumber
	&G^{ox}_y(x,y;x',y')= \\ \nonumber
	&\frac{2}{t_{ox}}\sum_{n=1}^{\infty}\frac{sin\left(k_n^{II}\left(t_1-y\right)\right)sin\left(k_n^{II}\left(t_1-y'\right)\right)}{k_n^{II}sinh\left(k_n^{II}L\right)}\times\;\;\;\;\;\;\;\;\;\;\;\;\;\;\;\;\;\;\;\;\;\;\;\;\;\;\;\;\;\; \\ \nonumber
	&\big[sinh\left(k_n^{II}x\right)sinh\left(k_n^{II}\left(L-x'\right)\right)\;\;\;\;\; x<x' \\ 
	&sinh\left(k_n^{II}x'\right)sinh\left(k_n^{II}\left(L-x\right)\right)\;\;\;\;\; x>x\big]
\end{align}

The magnetic scalar functions are obtained by plugging Green's functions into the following Green's identity.   
\begin{widetext}
\begin{align}
	\phi (x,y)= \iint{\left(\nabla . M\right)}G(x,y;x',y')dx'dy'-\oint\frac{\partial G(x',y')} {\partial n'}\phi(x',y')dS'+\oint\frac{\partial\phi(x',y')}{\partial n'}G(x,y;x',y')dS' \label{Green}
\end{align}
Eq. (\ref{MD_pot}) contains a $\phi_{MD}(x,y)$ term, which caused by the multi-domain in the ferromagnetic layer. The first term in $\left(\ref{Green}\right)$ corresponds to the following potential component in (\ref{MD_pot}). 
\begin{align}
	&\phi_{MD}(x,y)=\frac{2M_s}{t_{fm}T_Dd_w}\sum_{n}\ \frac{sin \left(k_{n}^{I}y\right)}{k_{n}^{I}}\sum_{j}\left [ \left ( \frac{k_{n}^{I}}{\left ( k_{n}^{I} \right )^2+k_j^2} \right )\left(\frac{1}{k_j^2}\right)\left \{ B_j\left ( \lambda_0^nI_1^{j,n}k_j+\lambda_1^nI_2^{j,n}  \right ) -A_j\left ( \lambda_0^nI_2^{j,n}k_j-\lambda_1^nI_1^{j,n}  \right )\right \} \right ]
	\\
	&\lambda_0=\frac{2(-1)^{n}}{t_{fm}\left ( k_{n}^{I} \right )^2}+\frac{1}{k_{n}^{I}}\;;\lambda_1=\frac{2(-1)^{n}}{t_{fm}^2\left ( k_{n}^{I} \right )^2}\\
	&I_1^{j,n} =cos(k_jx)-\frac{sinh\left ( k_{n}^{I}\left ( L-x \right ) \right )}{sinh\left ( k_{n}^{I}L \right )}-\frac{cos(k_jL)sinh\left ( k_{n}^{I}x \right )}{sinh\left ( k_{n}^{I}L \right )}\\ &I_2^{j,n}=sin(k_jx)-\frac{sin(k_jL)sinh\left ( k_{n}^{I}x \right )}{sinh\left ( k_{n}^{I}L \right )} \\	
	&A_j=cos\left ( k_jd_w \right )-1+cos\left ( k_j\left ( x_1+d_w \right ) \right )-cos\left ( k_j\left ( x_1+3d_w \right ) \right )+cos\left ( k_j\left ( x_1+x_2+4d_w \right ) \right )-+cos\left ( k_j\left ( x_1+x_2+3d_w \right ) \right )\\
	&B_j=sin\left ( k_jd_w \right )+sin\left ( k_j\left ( x_1+d_w \right ) \right )-sin\left ( k_j\left ( x_1+3d_w \right ) \right )+sin\left ( k_j\left ( x_1+x_2+4d_w \right ) \right )-sin\left ( k_j\left ( x_1+x_2+3d_w \right ) \right )
\end{align}
$B_s^m$ in $\left(\ref{MD_pot}\right)$ signifies the Fourier coefficient associated with magnetic field vector.
 \begin{align}
 	B_s^m=\int_{0}^{L}B_s(x',t_{fm})sin\left(k_mx'\right)dx'
 \end{align}  
Fourier coefficient $B_s^m$ is calculated by the potential continuity boundary condition at y=$t_{fm}$ 
\begin{align}
	\phi_{fm}(x,t_{fm})=\phi_{ox}(x,t_{fm})
\end{align}
\begin{align}
		&B_s^m=\frac{d_1^m+d_2^{m,n}+d_{MD}^{j,m,n}-d_3^{m,n}-d_4^m}{d_5^m}\\[0.5em]
	&	d_1^m=\frac{H_0L\left(1+\left(-1\right)^{m+1}\right)}{k_m cosh\left(k_m t_{fm}\right)}\\[0.5em]
		&d_2^{m,n}=\frac{2}{t_{fm}}\sum_{n}\frac{sin\left(k_n^It_{fm}\right)k_m\left(A_1^n+A_2^n\left(-1\right)^{m+1}\right)}{k_m^2+\left(k_n^I\right)^2}\\[0.5em]
		&d_3^{m,n}=\frac{2}{t_{ox}}\sum_{n}\frac{sin\left(k_n^{II}t_{ox}\right)k_m\left(B_1^n+B_2^n\left(-1\right)^{m+1}\right)}{k_m^2+\left(k_n^{II}\right)^2}\\[0.5em]
	&	d_4^m=\frac{H_0L\left(1+\left(-1\right)^{m+1}\right)}{k_m cosh\left(k_m t_{ox}\right)}\\[0.5em]
		&d_5^m=\frac{1}{\mu_{fe}k_mtanh\left(k_mt_{fm}\right)}+\frac{tanh\left(k_mt_{ox}\right)}{k_m\mu_{ox}}	\\ \nonumber
		&d_{MD}^{j,m,n}=\\ 
		&\frac{2M_s}{t_{fm}T_Dd_w}\sum_{n}\ \frac{sin \left(k_{n}^{I}t_{fm}\right)}{k_{n}^{I}}\sum_{j}\left [ \left ( \frac{k_{n}^{I}}{\left ( k_{n}^{I} \right )^2+k_j^2} \right )\left(\frac{1}{k_j^2}\right)\left \{ B_j\left ( \lambda_0^nI_1^{j,m,n}k_j+\lambda_1^nI_2^{j,m,n}  \right ) -A_j\left ( \lambda_0^nI_2^{j,m,n}k_j-\lambda_1^nI_1^{j,m,n}  \right )\right \} \right ]\\
		&I_1^{j,m,n}= \left ( 1+(-1)^{m+1}cos(k_jL) \right )\left ( \frac{k_m}{k_m^2-k_j^2}-\frac{k_m}{\left (k_{n}^{I} \right )^2+k_m^2}  \right )
\end{align}
\begin{align}	
	&I_2^{j,m,n}= \left ( (-1)^{m+1}sin(k_jL) \right )\left ( \frac{k_m}{k_m^2-k_j^2}-\frac{k_m}{\left (k_{n}^{I} \right )^2+k_m^2}  \right )
\end{align}
Fourier series coefficients at left (x=0), and right (x=L) are calculated by assuming a linear potential profile at these boundary gaps.
\begin{align}
	\phi_{fm}(0,y)=a_1y+H_0 \label{11}\\
	\phi_{ox}(0,y)=b_1y+b_2 \label{22}
\end{align}
where, $H_0$ is the externally applied uniform magnetic field. The unknown potential constants in $\left(\ref{11}\right)$ and $\left(\ref{22}\right)$ are calculated by the following boundary conditions.
\begin{align}
&\phi_{fm}(0,t_{fm})=\phi_{ox}(0,t_{fm})\\
&\phi_{ox}(0,t_{1})=H_0\\
&-\mu_{0} \frac{\partial \phi_{fm}(0,t_{fm})}{\partial y}+\mu_0 M(0,t_{fm})=\mu_{ox} \frac{\partial \phi_{ox}(0,t_{fm})}{\partial y}
\\
&A_1^n=\int_{0}^{t_{fm}}sin\left(k_{n}^{I}y\right)\phi_{fm}(0,y)dy\\
&B_1^n=\int_{0}^{t_{fm}}sin\left (k_{n}^{II}\left(t_{fm}+t_{ox}-y\right)  \right )\phi_{ox}(0,y)dy
\end{align}
$A_2^n$ and $B_2^n$ are calculated in the same manner by replacing $M(0,t_{fm})$ with $M(L,t_{fm})$.\\
Default Parameters: $t_{fm}$ = 20 nm, $t_{ox}$ =  4 nm, L =5$\times$$T_D$, $t_1=t_{fm}+t_{ox}$, $k_m=m\pi/L$, $k_n^{I}=(2n-1)\pi/(2t_{fm})$, $k_n^{II}=(2n-1)\pi/(2t_{ox})$, $\mu_{fe}$=5$\times$10$^3 \mu_{0}$, $\mu_{ox}$=1$\times$$\mu_{0}$, $k_u$ = 5$\times$10$^{-4}$ J/m$^3$, $M_s$=6$\times$10$^{5}$A/m, $A$=1$\times$10$^{-11}$ J/m, $D$= 1$\times$10$^{-3}$ J/m$^2$, $\theta = \pi/2$
\end{widetext}
\end{ceqn}

\end{document}